\documentclass[sigplan,11pt,nonacm,natbib=false]{acmart}
\usepackage[maxnames=1,minnames=1,maxbibnames=100,natbib=true,citestyle=authoryear,bibstyle=authoryear,doi=false,url=false,isbn=false,isbn=false,backend=biber]{biblatex}
\ifnum\pdfshellescape=1
  \usepackage[finalizecache]{minted}
\else
  \usepackage[frozencache]{minted}
\fi
\usepackage[nocn]{ffcode}
\usepackage{tikz}
\usepackage{to-be-determined}
\title{Reducing Programs to Objects}
\subtitle{
  Ver:
  \texorpdfstring{
    \href{https://github.com/objectionary/reducing-programs-to-objects/releases/tag/0.2.0}
      {\ff{0.2.0}}
  }{0.2.0}
}
\author{Yegor Bugayenko}
\affiliation{\institution{Huawei}\country{Russia}\city{Moscow}}
\email{yegor256@gmail.com}

\begin{abstract}
C++, Java, C\#, Python, Ruby, JavaScript are the most powerful object-oriented programming languages, if language power would be defined as the number of \emph{features} available for a programmer. EO, on the other hand, is an object-oriented programming language with a reduced set of features: it has nothing by objects and mechanisms of their composition and decoration. We are trying to answer the following research question: ``Which known features are possible to implement using only objects?''
\end{abstract}

\begin{document}
\raggedbottom

\maketitle

\section{Features}

To answer our research question we selected most complex features and demonstrated how each of them may be represented in EO~\citep{bugayenko2021eolang} objects\footnote{%
\LaTeX{} sources of this paper are maintained in
\href{https://github.com/objectionary/reducing-programs-to-objects}{REPOSITORY} GitHub repository,
the rendered version is \href{https://github.com/objectionary/reducing-programs-to-objects/releases/tag/0.2.0}{\ff{0.0.0}}.}:

\begin{itemize}
    \item Non-conditional jumps
      (Sec.~\ref{sec:goto}),
    \item Data and code pointers
      (Sec.~\ref{sec:pointers}),
    \item Procedures
      (Sec.~\ref{sec:procedures}),
    \item Classes
        (Sec.~\ref{sec:classes}),
    \item Exceptions
        (Sec.~\ref{sec:exceptions}),
    \item Anonymous functions
        (Sec.~\ref{sec:blocks}),
    \item Generators
        (Sec.~\ref{sec:generators}),
    \item Types and casting
        (Sec.~\ref{sec:types}),
    \item Reflection
        (Sec.~\ref{sec:reflection}),
    \item Static methods
        (Sec.~\ref{sec:static}),
    \item Inheritance
        (Sec.~\ref{sec:inheritance}),
    \item Method overloading
        (Sec.~\ref{sec:overloading}),
    \item Java generics
        (Sec.~\ref{sec:generics}),
    \item C++ templates
        (Sec.~\ref{sec:templates}),
    \item Mixins
        (Sec.~\ref{sec:mixins}),
    \item Java annotations
        (Sec.~\ref{sec:annotations}).
\end{itemize}

Other features are more trivial, that's why they are not presented in this paper, such as operators, loops, variables, code blocks, constants, branching, and so on.

\subsection{Goto}
\label{sec:goto}

Goto is a one-way imperative transfer of control to another line of code. There are only two possible cases of goto jumps: forward and backward.

\subsubsection{Backward Jump}

This is a ``backward'' jump example in C language:

\begin{ffcode}
#include "stdio.h"
void f() {
  int i = 1;
  again:
  i++;
  if (i < 10) goto again;
  printf("Finished!");
}
\end{ffcode}

It can be mapped to the following EO code:

\begin{ffcode}
[] > f
  memory 0 > i
  seq > @
    i.write 1
    goto
      [g]
        seq > @
          i.write (i.plus 1)
          if.
            i.lt 10
            g.backward
            TRUE
    QQ.io.stdout "Finished!"
\end{ffcode}

Here, the one-argument abstract atom \ff{goto} is being copied with a one-argument abstract anonymous object, which is the sequence of objects doing the increment of \ff{i} and then the comparison of it with the number ten. If the condition is true, \ff{g.backward} is called, which leads to a backward jump and re-iteration of \ff{goto}.

\subsubsection{Forward Jump}

This is an example of a ``forward'' jump in C language:

\begin{ffcode}
int f(int x) {
  int r = 0;
  if (x == 0) goto exit;
  r = 42 / x;
  exit:
  return r;
}
\end{ffcode}

It can be mapped to the following EO code:

\begin{ffcode}
[x] > f
  memory 0 > r
  seq > @
    r.write 0
    goto
      [g]
        seq > @
          if.
            x.eq 0
            g.forward TRUE
            TRUE
          r.write (42.div x)
    r
\end{ffcode}

Here, the same abstract atom \ff{goto} is copied with an abstract one-argument object, which is a sequence of objects. When the condition is true, a forward jump is performed by \ff{g.forward} atom.

Similar way, the atom \ff{goto} may be used to simulate other conditional jump statements, like \ff{break}, \ff{continue}, or \ff{return} in the middle of a function body (see Sec.~\ref{sec:procedures}).

\subsubsection{Complex Case}

This is a more complex case of using \ff{goto} in C:

\begin{ffcode}
#include "stdio.h"
void f(int a, int b) {
  goto start;
back:
  printf("A");
start:
  if (a > b) goto back;
  printf("B");
}
\end{ffcode}

\begin{figure}
\includegraphics[width=0.9\columnwidth]{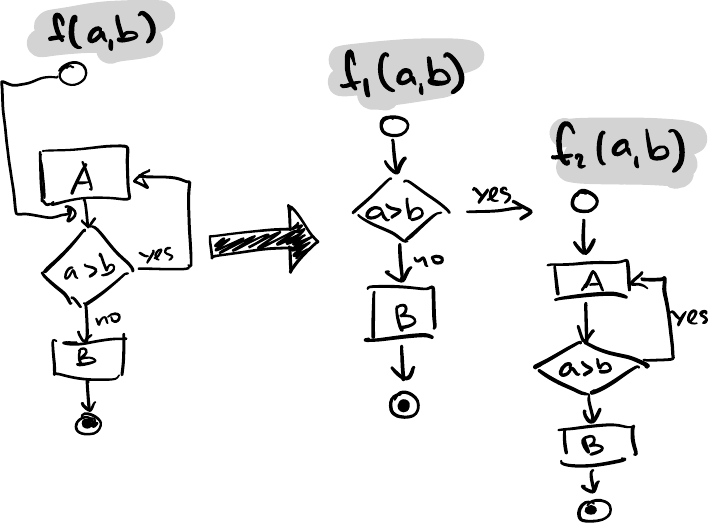}
\caption{The function \ff{f} with a few \ff{goto} statements inside is translated to two functions, reducing the complexity of the code at the cost of introducing duplication.}
\Description{Workflow of GOTO object}
\label{fig:goto}
\end{figure}

In order to translate this code to EO it has to be refactored as Fig.~\ref{fig:goto} demonstrates. The function \ff{f()} is copied twice and each copy has its own execution flow implemented.

\begin{ffcode}
#include "stdio.h"
void f(int a, int b) { f1(a, b); }
void f1(int a, int b) {
  if (a > b) f2(a, b);
  printf("B");
}
void f2(int a, int b) {
back:
  printf("A");
  if (a > b) goto back;
  printf("B");
}
\end{ffcode}

Then, the translation to EO is trivial, with the use of the \ff{goto} object.

In more complex cases a program may first be restructured to replace \ff{goto} statements with loops and branching, as suggested by~\citet{williams1985restructuring,pan1996formal,erosa1994taming,ceccato2008goto}.

\subsubsection{Multiple Returns}

Some structured programming languages allow a function to exit at an arbitrary place, not only at the end, using \ff{return} statements, for example:

\begin{ffcode}
void abs(int x) {
  if (x > 0) {
    return x;
  }
  return -1 * x;
}
\end{ffcode}

This can be mapped to the following EO code using \ff{goto} object:

\begin{ffcode}
[x] > abs
  goto > @
    [g]
      seq > @
        if.
          x.gt 0
          g.forward x
          TRUE
        g.forward
          -1.times x
\end{ffcode}

The dataization of \ff{g.forward} will exit the \ff{goto} object wrapping the entire code in the ``function'' \ff{abs}.

\subsection{Pointers}
\label{sec:pointers}

A pointer is an object in many programming languages that stores a memory address. Pointers may point to data memory and to program memory.

\subsubsection{Pointers to Data}

This is an example of C code, where the pointer \ff{p} is first incremented by seven times \ff{sizeof(book)} (which is equal to 108) and then de-referenced to become a struct \ff{book} mapped to memory. Then, the \ff{title} part of the struct is filled with a string and the \ff{price} part is returned as a result of the function \ff{f}:

\begin{ffcode}
#include "string.h"
struct book {
  char title[100];
  long long price;
};
int f(struct book* p) {
  struct book b = *(p + 7);
  strcpy(b.title, "Object Thinking");
  return b.price;
}
\end{ffcode}

This code can be mapped to the following EO code:

\begin{ffcode}
[ptr] > book
  ptr > @
  ptr.block > title
    100
    [b] (b.as-string > @)
  ptr.block > price
    8
    [b] (b.as-int > @)
pointer. > p1
  heap 1024
  0
  108
[p] > f
  seq > @
    book (p.add 7) > b
    b.title.write
      ("Object Thinking").as-bytes
    b.price
\end{ffcode}

Here, \ff{p1} is an object that can be used as an argument of \ff{f}. It is a copy of an abstract object \ff{heap.pointer}, which expects two: 1)~an absolute address in memory, and 2)~the size of the memory block it points to, in bytes. Its attribute \ff{block} expects two attributes: 1) the number of bytes in the heap, and 2)~an abstract object that can encapsulate \ff{bytes} which were just read from memory.

The object \ff{heap} is an abstraction of a random access memory.

\subsubsection{Pointers to Code}

A pointer may not only refer to data in the heap, but also to executable code in memory (there is no difference between ``program'' and ``data'' memories both in x86 and ARM architectures). This is an example of C program, which calls a function referenced by a function pointer, providing two integer arguments to it:

\begin{ffcode}
int foo(int x, int y) {
  return x + y;
}
int f() {
  int (*p)(int, int);
  p = &foo;
  return (*p) (7, 42);
}
\end{ffcode}

This code can be mapped to the following EO code:

\begin{ffcode}
[x y] > foo
  x.plus y > @
[] > f
  cage 0 > p
  seq > @
    p.write
      [x y]
        foo x y > @
    p.@ 7 42
\end{ffcode}

Important to notice that the following code is not possible to represent in EO:

\begin{ffcode}
int f() {
  int (*p)(int);
  p = (int(*)(int)) 0x761AFE65;
  return (*p) (42);
}
\end{ffcode}

Here, the pointer refers to an arbitrary address in program memory, where some executable code is supposed to be located.

\subsubsection{Variables in Stack}

This C function, which returns seven (tested with GCC), assumes that variables are placed in stack sequentially and uses pointer de-referencing to access them:

\begin{ffcode}
int f() {
  long long a = 42;
  long long b = 7;
  return *(&a - 1);
}
\end{ffcode}

This code can be mapped to the following EO code:

\begin{ffcode}
[p] > long64
  p.block > @
    8
    [b] (b.as-int > @)
[] > f
  seq > @
    malloc. > stack
      heap 32 > h
      16
    long64 stack > b
    b.write (7.as-bytes)
    long64 stack > a
    a.write (42.as-bytes)
    long64 (a.p.sub 1) > ret!
    h.free stack
    ret
\end{ffcode}

Here, the atom \ff{malloc} allocates a segment of bytes in the \ff{heap}. Later, the atom \ff{free} releases it. The attribute \ff{ret} is made constant in order to avoid its recalculation after it's ``returned'' (this mechanism is explained in Sec.~\ref{sec:destructors} where destructors are discussed).

\subsection{Procedures}
\label{sec:procedures}

A subroutine is a sequence of program instructions that performs a specific task, packaged as a unit. In different programming languages, a subroutine may be called a routine, subprogram, function, method, or procedure. In this PHP example, a function \ff{max} is defined:

\begin{ffcode}
function max($a, $b) {
  if ($a > $b) return $a;
  return $b;
}
\end{ffcode}

It can be mapped to the following EO code:

\begin{ffcode}
[a b] > max
  goto > @
    [g]
      seq > @
        if.
          a.gt b
          g.forward a
          TRUE
        b
\end{ffcode}

This example also demonstrates how \ff{goto} object can be used to simulate the behavior of the \ff{return} statement from within the body of a method.

\subsubsection{Impure Functions}

A function is pure when, among other qualities, it doesn't have side effects:
no mutation of local static variables, non-local variables, mutable reference arguments
or input/output streams. This PHP function is impure:

\begin{ffcode}
$a = 42;
function inc($a) {
  $a = $a + 1;
  return $a;
}
inc(inc($a)); // $a == 44
\end{ffcode}

It may be mapped to a constant EO object:

\begin{ffcode}
memory 42 > a
[x] > inc!
  seq > @
    x.write
      x.plus 1
inc
  inc a
\end{ffcode}

\subsection{Classes}
\label{sec:classes}

A class is an extensible program-code-template for creating objects, providing initial values for state (member variables) and implementations of behavior (member functions or methods). The following program contains a Ruby class with a single constructor, two methods, and one mutable member variable:

\begin{ffcode}
class Book
  def initialize(i)
    @id = id
    puts "New book!"
  end
  def path
    "/tmp/${@id}.txt"
  end
  def move(i)
    @id = i
  end
end
\end{ffcode}

It can be mapped to the following EO code, where a class becomes a factory of objects:

\begin{ffcode}
[i] > book
  cage i > id
  [] > ruby-init
    QQ.txt.sprintf > @
      "New book!"
  [] > path
    QQ.txt.sprintf > @
      "/tmp/
      id
  [i] > move
    id.write i > @
\end{ffcode}

Here, the constructor is represented by the object \ff{ruby-init}, which initializes finishes the initization of the object. The making an ``instance'' of a book would look like this:

\begin{ffcode}
book 42 > b
b.ruby-init
b.move 7
b.path
\end{ffcode}

\subsection{Destructors}
\label{sec:destructors}

In C++ and some other languages, a destructor is a method that is called for a class object when that object passes out of scope or is explicitly deleted. The following code will print both \ff{Alive} and \ff{Dead} texts:

\begin{ffcode}
#include <iostream>
class Foo {
public:
  Foo() { std::cout << "Alive"; }
  ~Foo() { std::cout << "Dead"; };
};
int main() {
  Foo f = Foo();
}
\end{ffcode}

It may be translated to EO as such:

\begin{ffcode}
[] > foo
  [] > constructor
    QQ.io.stdout "Alive" > @
  [] > destructor
    QQ.io.stdout "Dead" > @
[] > main
  foo > f
  seq > @
    f.constructor
    f.destructor
\end{ffcode}

There is no garbage collection in EO, that's why a destructor must be explicitly ``called'' when an object passes out of scope or is deleted.

\subsection{Exceptions}
\label{sec:exceptions}

Exception handling is the process of responding to the occurrence of exceptions---anomalous or exceptional conditions requiring special processing---during the execution of a program provided. This C++ program utilizes exception handling to avoid segmentation fault due to null pointer de-referencing:

\begin{ffcode}
#include <iostream>
class Book { public: int price(); };
int price(Book* b) {
  if (b == NULL) throw "NPE!";
  return (*b).price() * 1.1;
}
void print(Book* b) {
  try {
    std::cout << "The price: " << price(b);
  } catch (char const* e) {
    std::cout << "Error: " << e;
  }
}
\end{ffcode}

This mechanism may be implemented in EO:

\begin{ffcode}
[] > book
  [] > price /int
[b] > price
  if. > @
    b.eq 0
    error "NPE!"
    b.price.times 1.1
[b] > print
  try > @
    []
      QQ.io.stdout > @
        QQ.txt.sprintf
        "The price: 
        price b
    [e]
      QQ.io.stdout > @
        QQ.txt.sprintf
        "Error: 
        e
    []
      nop > @
\end{ffcode}

Here, the object \ff{try} expects three arguments: 1)~an abstract object to be dataized, and 2)~an abstract object to be copied with one argument, in case dataization returns encapsulated object, and 3)~an object to be dataized anyway (similar to Java \ff{finaly} block).

In the object \ff{price} we get the object \ff{error}, which if dataized, causes the termination of dataization and a non-conditional jump to the ``catch'' object of the \ff{try}. The mechanism is similar to ``checked'' exceptions in Java, where a method's signature must declare all types of exceptions the method may throw.

\subsubsection{Many Exception Types}

A Java method may throw a number of either checked or unchecked exceptions, for example:

\begin{ffcode}
void f(int x) throws IOException {
  if (x == 0) {
    throw new IOException();
  }
  throw new RuntimeException();
}
\end{ffcode}

This would be represented in EO as such:

\begin{ffcode}
[x] > f
  if. > @
    x.eq 0
    error "IOException"
    error "RuntimeException"
\end{ffcode}

To catch both exceptions the object \ff{f} would be used like this:

\begin{ffcode}
try
  []
    try > @
      []
        f 5 > @
      [e1]
        QQ.io.stdout e1 > @
      []
        nop > @
  [e2]
    QQ.io.stdout e2 > @
  []
    nop > @
\end{ffcode}

\subsection{Anonymous Functions}
\label{sec:blocks}

Anonymous functions that can be passed into methods as arguments. For example, in this Ruby code a ``block'' (a name Ruby uses for anonymous functions) is passed:

\begin{ffcode}
def scan(lines)
  lines.each do |t|
    if t.starts_with? '#' yield t
  end
end
scan(array) { |x| puts x }
\end{ffcode}

This mechanism may be implemented in EO:

\begin{ffcode}
[lines b] > scan
  lines.each > @
    [t]
      if. > @
        t.starts-with "#"
        b t
        TRUE
scan
  array
  [x]
    QQ.io.stdout x > @
\end{ffcode}

Here, the anonymous function passed to the object \ff{scan} as an argument \ff{b}. The ``call'' of this function is the dataization of its copy, performed by the \ff{if} atom.

\subsection{Generators}
\label{sec:generators}

A generator is a routine that can be used to control the iteration behaviour of a loop. For example, this PHP program will print first ten Fibonacci numbers:

\begin{ffcode}
function fibonacci(int $limit):generator {
  yield $a = $b = $i = 1;
  while (++$i < $limit) {
    $b = $a + $b;
    yield $b - $a;
    $a = $b;
  }
}
foreach (fibonacci(10) as $n) {
  echo "$n\n";
}
\end{ffcode}

This mechanism may be implemented in EO:

\begin{ffcode}
[limit f] > fibonacci
  memory 0 > a
  memory 0 > b
  memory 0 > i
  seq > @
    a.write 1
    b.write 1
    f 0
    while.
      seq (i.write (i.plus 1)) (i.lt limit)
      [idx]
        seq > @
          b.write (a.plus b)
          a.write (b.minus a)
          f (b.minus a)
    TRUE
fibonacci > @
  10
  [n]
    QQ.io.stdout > @
      QQ.txt.sprintf "
\end{ffcode}

Here, the generator is turned into an abstract object \ff{fibonacci} with an extra parameter \ff{f}, which is specified with an abstract object argument, which prints what's given as \ff{n}.

\subsection{Types and Type Casting}
\label{sec:types}

A type system is a logical system comprising a set of rules that assigns a property called a type to the various constructs of a computer program, such as variables, expressions, functions or modules. The main purpose of a type system is to reduce possibilities for bugs in computer programs by defining interfaces between different parts of a computer program, and then checking that the parts have been connected in a consistent way. In this example, a Java method \ff{add} expects an argument of type \ff{Book} and compilation would fail if another type is provided:

\begin{ffcode}
class Cart {
  private int total;
  void add(Book b) {
    this.total += b.price();
  }
}
\end{ffcode}

The restrictions enforced by Java type system in compile time through types \ff{Cart} and \ff{Book} may be represented in EO by means of decorators, for example:

\begin{ffcode}
[] > original-cart
  memory 0 > total
  [b] > add
    total.write (total.plus (b.price))
[] > cart
  original-cart > @
  [b] > add
    if.
      b.subtype-of (QQ.txt.text "Book")
      @.add b
      []
        error "Type mismatch, Book expected" > @
\end{ffcode}

Here, it is expected that the parameter \ff{b} is defined in a ``class,'' which has \ff{subtype-of} attribute, which may also be provided by a decorator (a simplified example):

\begin{ffcode}
[] > original-book
  memory 0 > price
[] > book
  original-book > @
  [t] > subtype-of
    t.eq "Book" > @
  [] > price
    @.price > @
\end{ffcode}

This decoration may be simplified through a ``fluent'' supplementary object:

\begin{ffcode}
type "Cart" > cart-type
.super-types "Object" "Printable"
.method "add" "Book"
cart-type original-cart > cart
\end{ffcode}

Here, the \ff{type} object implements all necessary restrictions Java type system may provide for the type \ff{Cart} and its methods.

Type casting, which is a mechanism of changing an expression from one data type to another, may also be implemented through same decorators.

\subsection{Reflection}
\label{sec:reflection}

Reflection is the ability of a process to examine, introspect, and modify its own structure and behavior. In the following Python example the method \ff{hello} of an instance of class \ff{Foo} is not called directly on the object, but is first retrieved through reflection functions \ff{globals} and \ff{getattr}, and then executed:

\begin{ffcode}
def Foo():
    def hello(self, name):
        print("Hello, 
obj = globals()["Foo"]()
getattr(obj, "hello")("Jeff")
\end{ffcode}

It may be implemented in EO just by encapsulating additional meta information in classes explained in Sec.~\ref{sec:classes}.

\subsubsection{Monkey Patching}

Monkey patching is making changes to a module or a class while the program is running. This JavaScript program adds a new method \ff{print} to the object \ff{b} after the object has already been instantiated:

\begin{ffcode}
function Book(t) { this.title = t; }
var b = new Book("Object Thinking");
b.print = function() {
  console.log(this.title);
}
b.print();
\end{ffcode}

This program may be translated to EO, assuming that \ff{b} is being held by an attribute of a larger object, for example \ff{app}:

\begin{ffcode}
[] > app
  cage 0 > b
  b' > copy
  seq > @
    b.write
      [] > book
        [t] > new
          memory "" > title
          title.write t > @
    copy.<
    b.write
      [] > book
        [t] > new
          seq > @
            []
              copy > @
              [] > print
                QQ.io.stdout (^.title)
    b.print
\end{ffcode}

Here, the modification to the object \ff{book} is happening through making a copy of it, creating a decorator, and then storing it to where the original object was located.

\subsection{Static Methods}
\label{sec:static}

A static method (or static function) is a method defined as a member of an object but is accessible directly from an API object's constructor, rather than from an object instance created via the constructor. For example, this C\# class consists of a single static method:

\begin{ffcode}
class Utils {
  public static int max(int a, int b) {
    if (a > b) return a;
    return b;
  }
}
\end{ffcode}

It may be converted to the following EO code, since a static method is nothing else but a ``global'' function:

\begin{ffcode}
[a b] > utils-max
  if. > @
    a.gt b
    a
    b
\end{ffcode}

\subsection{Inheritance}
\label{sec:inheritance}

Inheritance is the mechanism of basing a class upon another class retaining similar implementation. In this Java code class \ff{Book} inherits all methods and non-private attributes from class \ff{Item}:

\begin{ffcode}
class Item {
  private int p;
  int price() { return p; }
}
class Book extends Item {
  int tax() { return price() * 0.1; }
}
\end{ffcode}

It may be represented in EO like this:

\begin{ffcode}
[] > item
  memory 0 > p
  [] > price
    p > @
[] > book
  item > i
  [] > tax
    (QQ.math.number (i.price)).as-float.times 0.1 > @
\end{ffcode}

Here, composition is used instead of inheritance.

\subsubsection{Prototype-Based Inheritance}

So called pro\-to\-type-based programming uses generalized objects, which can then be cloned and extended. For example, this JavaScript program defines two objects, where \ff{Item} is the parent object and \ff{Book} is the child that inherits the parent through its prototype:

\begin{ffcode}
function Item(p) { this.price = p; }
function Book(p) {
  Item.call(this, p);
  this.tax = function () {
    return this.price * 0.1;
  }
}
var t = new Book(42).tax();
console.log(t); // prints "4.2"
\end{ffcode}

This mechanism of prototype-based inheritance may be translated to the following EO code, using the mechanism of decoration:

\begin{ffcode}
[p] > item
  memory 0 > price
  [] > new
    price.write p > @
[] > book
  [p] > new
    item p > @
    [] > tax
      times. > @
        as-float.
          QQ.math.number (^.price)
        0.1
QQ.io.stdout
  QQ.txt.sprintf
    "
    tax.
      book.new 42
\end{ffcode}

\subsubsection{Multiple Inheritance}

Multiple inheritance is a feature of some object-oriented computer programming languages in which an object or class can inherit features from more than one parent object or parent class. In this C++ example,
the class \ff{Jack} has both \ff{bark} and \ff{listen} methods, inherited from \ff{Dog} and \ff{Friend} respectively:

\begin{ffcode}
#include <iostream>
class Dog {
  virtual void bark() {
    std::cout << "Bark!";
  }
};
class Friend {
  virtual void listen();
};
class Jack: Dog, Friend {
  void listen() override {
    Dog::bark();
    std::cout << "Listen!";
  }
};
\end{ffcode}

It may be represented in EO like this:

\begin{ffcode}
[] > dog
  [] > bark
    QQ.io.stdout "Bark!" > @
[] > friend
  [] > listen
[] > jack
  dog > d
  friend > f
  [] > listen
    seq > @
      d.bark
      QQ.io.stdout "listen!"
\end{ffcode}

Here, inherited methods are explicitly listen as attributes in the object \ff{jack}. This is very close to what would happen in the virtual table of a class \ff{Jack} in C++. The EO object \ff{jack} just makes it explicit.

\subsection{Method Overloading}
\label{sec:overloading}

Method overloading is the ability to create multiple functions of the same name with different implementations. Calls to an overloaded function will run a specific implementation of that function appropriate to the context of the call, allowing one function call to perform different tasks depending on context. In this Kotlin program two functions are defined with the same name, while only one of them is called with an integer argument:

\begin{ffcode}
fun foo(a: Int) {}
fun foo(a: Double) {}
foo(42)
\end{ffcode}

It may be represented in EO like this:

\begin{ffcode}
[args...] > foo
  (args.at 0) > a0
  if.
    a0.subtype-of "Int"
    first-foo a0
    second-foo a0
foo 42
\end{ffcode}

This code expects arguments of \ff{foo} to be equipped with the type system suggested in Sec.~\ref{sec:types}. The attribute \ff{subtype-of} will help dispatching the call to the right objects.

\subsection{Java Generics}
\label{sec:generics}

Generics extend Java's type system to allow a type or method to operate on objects of various types while providing compile-time type safety. For example, this Java class expects another class to be specified as \ff{T} before usage:

\begin{ffcode}
class Cart<T extends Item> {
  private int total;
  void add(T i) {
    total += i.price();
  }
}
\end{ffcode}

It may be represented in EO like this:

\begin{ffcode}
[] > cart
  memory 0 > total
  [i] > add
    total.write > @
      total.plus (i.price)
\end{ffcode}

As the example demonstrates, the presence of generics in class declaration may be ignored, since EO is a language without types and type checking.

\subsection{C++ Templates}
\label{sec:templates}

Templates are a feature of the C++ programming language that allows functions and classes to operate with generic types, allowing a function or class to work on many different data types without being rewritten for each one.

\begin{ffcode}
template<typename T> T max(T a, T b) {
  return a > b ? a : b;
}
int x = max(7, 42);
\end{ffcode}

It may be represented in EO like this:

\begin{ffcode}
[a b] > max
  if. > @
    a.gt b
    a
    b
max 7 42 > x
\end{ffcode}

As the example demonstrates, the presence of templates may be ignored, since EO is a language without types and type checking.

\subsection{Mixins}
\label{sec:mixins}

A mixin is a class that contains methods for use by other classes without having to be the parent class of those other classes. The following code demonstrates how Ruby module is included into a class:

\begin{ffcode}
module Timing
  def recent?
    @time - Time.now < 24 * 60 * 60
  end
end
def News
  include Timing
  def initialize(t)
    @time = t
  end
end
n = News.new(Time.now)
n.recent?
\end{ffcode}

This code may be represented in EO by just copying the method \ff{recent?} to the object \ff{News} as if it was defined there.

\subsection{Annotations}
\label{sec:annotations}

In Java, an annotation is a form of syntactic metadata that can be added to source code; classes, methods, variables, parameters and Java packages may be annotated. Later, annotations may be retrieved through Reflection API. For example, this program analyzes the annotation attached to the class of the provided object and changes the behavior depending on one of its attributes:

\begin{ffcode}
interface Item {}
@Ship(true) class Book { /*..*/ }
@Ship(false) class Song { /*..*/ }
class Cart {
  void add(Item i) {
    /* add the item to the cart */
    if (i.getClass()
      .getAnnotation(Ship.class)
      .value()) {
      /* mark the cart as shippable */
    }
  }
}
\end{ffcode}

The code may be represented in EO using classes suggested in Sec.~\ref{sec:classes}:

\begin{ffcode}
[] > book
  ship TRUE > a1
  [] > new
    [] > @
[] > song
  ship FALSE > a1
  [] > new
    [] > @
[] > cart
  [i] > add
    # Add the item to the cart
    if. > @
      i.a1.value
        # Mark the cart shippable
        TRUE
        FALSE
\end{ffcode}

Annotations become attributes of objects, which represent classes in EO, as explained in Sec.~\ref{sec:classes}: \ff{book.a1} and \ff{song.a1}.

\section{Traceability}

Certain amount of semantic information may be lost during the translation from a more powerful programming language to EO objects, such as, for example, namings, line numbers, comments, etc. Moreover, it may be useful to have an ability to trace EO objects back to original language constructs, which they were motivated by. For example, a simple C function:

\begin{ffcode}
int f(int x) {
  return 42 / x;
}
\end{ffcode}

It may be mapped to the following EO object:

\begin{ffcode}
[x] > f
  [] > @
    "src/main.c:1-1" > source
    42.div x > @
  "src/main.c:0-2" > source
\end{ffcode}

Here, synthetic \ff{source} attribute represents the location in the source code file with the original C code. It's important to make sure during translation that the name \ff{source} doesn't conflict with a possibly similar name in the object.

\section{Conclusion}

We demonstrated how some language features often present in high-level object-oriented languages, such as Java or C++, may be expressed using objects. We used EO as a destination programming language because of its minimalistic semantics: it doesn't have any language features aside from objects and their composition and decoration mechanisms.

\printbibliography

\clearpage

\end{document}